\documentclass[a4paper,fleqn]{cas-dc}

\usepackage[authoryear]{natbib}
\usepackage{color,soul}
\usepackage[section]{placeins}
\usepackage{graphicx}
\def\tsc#1{\csdef{#1}{\textsc{\lowercase{#1}}\xspace}}
\tsc{WGM}
\tsc{QE}
\tsc{EP}
\tsc{PMS}
\tsc{BEC}
\tsc{DE}

\begin{document}
\let\WriteBookmarks\relax
\def\floatpagepagefraction{1}
\def\textpagefraction{.001}
\shorttitle{Improved Cell Barrier Characterization in Transwell Inserts by Electrical Impedance Spectroscopy}
\shortauthors{Linz et~al.}

\title [mode = title]{Cell Barrier Characterization in Transwell Inserts by Electrical Impedance Spectroscopy}                      



\author[1,2]{Georg Linz}


\address[1]{DWI - Leibniz Institute for Interactive Materials, Forckenbeckstr. 50, 52074 Aachen, Germany}

\author[1,2]{Suzana Djeljadini} 

\author[1,2]{Lea Steinbeck} 

\author[3]{Gurbet K\"ose} 
\author[3]{Fabian Kiessling} 
\author[1,2]{Matthias Wessling}

\address[2]{RWTH Aachen University, Aachener Verfahrenstechnik-Chemical Process Engineering, Forckenbeckstrasse 51, 52074, Aachen, Germany}

\address[3]{Institute for Experimental Molecular Imaging, Faculty of Medicine, RWTH Aachen University,
52074 Aachen, Germany}

\begin{abstract}
We describe an impedance-based method for cell barrier integrity testing. A four-electrode electrical impedance spectroscopy (EIS) setup can be realized by simply connecting a commercial chopstick-like electrode (STX-1) to a potentiostat allowing monitoring cell barriers cultivated in transwell inserts. Subsequent electric circuit modeling of the electrical impedance results the capacitive properties of the barrier next to the well-known transepithelial electrical resistance (TEER). The versatility of the new method was analyzed by the EIS analysis of a Caco-2 monolayer in response to (a) different membrane coating materials, (b) two different permeability enhancers ethylene glycol-bis(2-aminoethylether)-N,N,N',N'-tetraacetic acid (EGTA) and saponin, and (c) sonoporation. 
For the different membrane coating materials, the TEERs of the standard and new protocol coincide and increase during cultivation, while the capacitance shows a distinct maximum for three different surface materials (no coating, Matrigel\textsuperscript{\textregistered}, and collagen I). The permeability enhancers cause a decline in the TEER value, but only saponin alters the capacitance of the cell layer by two orders of magnitude. Hence, cell layer capacitance and TEER represent two independent properties characterizing the monolayer. The use of commercial chopstick-like electrodes to access the impedance of a barrier cultivated in transwell inserts enables remarkable insight into the behavior of the cellular barrier with no extra work for the researcher. This simple method could evolve into a standard protocol used in cell barrier research.
\end{abstract}


\maketitle

\section{Introduction}

Cell barriers form interfaces between tissue com\-part\-ments and control the exchange of substances. Tight cell contacts limit the unhindered diffusion between compartments and bring single cells to act as a barrier. Cell-cell connections are formed by different protein complexes, with the most prominent representative called tight junctions. 
They also regulate the paracellular transport and limit the diffusion of molecules and ions through intercellular spaces \citep{tsukita2001multifunctional}. These protein complexes enable the polarization of cells by different compositions of lipids and transmembrane proteins, leading to the formation of the apical and basolateral sides \citep{ebnet2015cell}. 
Passing a cell barrier is crucial for many biological processes and a major challenge in drug delivery \citep{benson2013impedance}.

Research has so far focused on developing the \textit{in-vitro} models of various cell barriers. Among others, barrier models for lungs \citep{hermanns2004lung}, intestinal \citep{hidalgo1989characterization}, and blood-brain \citep{czupalla2014vitro} have been described. A common method for \textit{in-vitro} experiments with barrier-forming cells is to cultivate the cells on a removable porous membrane support for a standard well plate. The removable insert allows access to the apical and basolateral sides~\citep{benson2013impedance}. 

Monitoring the development and integrity of cell barriers during maturation and experiments is crucial for all studies performed on barriers. Methods for accessing the barrier permeability are based on the transport of tracer substances, e.g., mannose or fluorescent dyes, immunofluorescent staining of proteins related to the tight junction complex, or measurement of the electrical resistance \citep{yeste2018engineering}. In this article, we focus on the latter considering that electric resistance is noninvasive, comparably easy to perform, and the most widely used method \citep{srinivasan2015teer}.

In particular, we focus on an important extension of the current methodology: by using the complex impedance of a TEER measurement more detailed information can be obtained about the various contributions to the barrier development.

\section{Background}

\subsection{TEER}
Electrical resistance rises from the high transport resistance of ions through a cell barrier and can, therefore, be used as a measure of the integrity of a barrier. The TEER is often measured with a chopstick-like electrode \citep{yeste2018engineering}.

\begin{figure*}
	\centering
		\includegraphics[width=2\columnwidth]{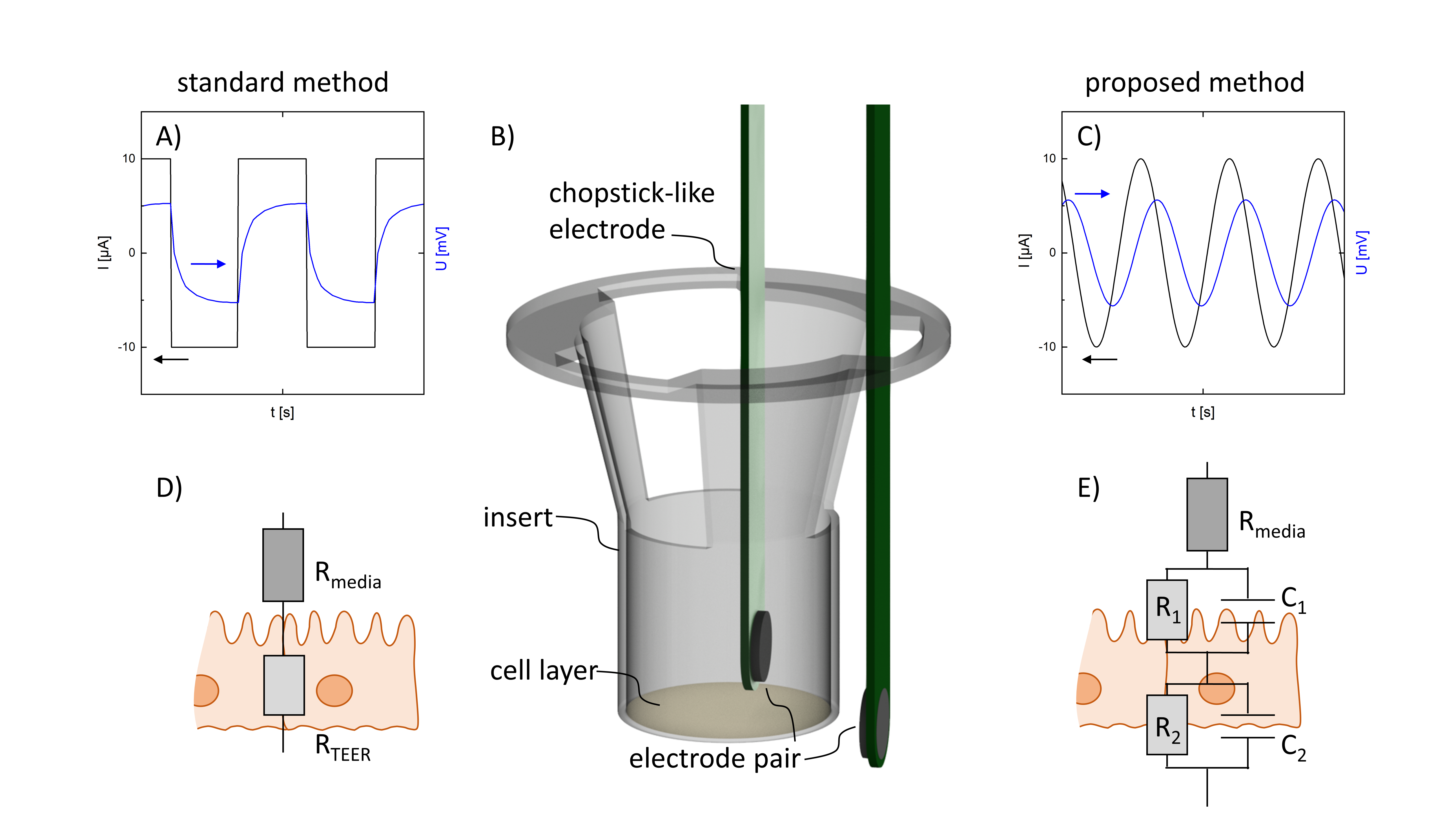}
	\caption{Graphic B displays the experimental setup of a chopstick-like electrode inserted in a transwell insert (B). When using the chopstick-like electrode with an endothelial voltohmmeter (standard method), the cell layer resistance is gained by applying a constant current and recording the resulting potential (A). The resistance can be calculated with Ohm's law and is the sum of the  media resistance $R_{Media}$ and the cell layer resistance $R_{TEER}$ (D). In the proposed method, the input signal is changed to an alternating current signal with varying frequencies (C). By equivalent electric circuit fitting cell layer resistances ($R_{1}$;$R_{2}$) and capacitances ($C_{1}$;$C_{2}$) are extracted from the obtained impedance data. The fitted impedance data leads to direct determination between media and cell layer resistance and additional cell layer capacitance data (E).}
\label{fig:messaufbau}
\end{figure*}

The standard method for measuring TEER is to connect the chopstick-like electrode with an
Epithelial voltohmmeter. The voltohmmeter imposes a square-wave current of 10~$\mu$A with 12.5 polarity reversals every second between the current-injecting electrodes. The two voltage pick-up electrodes on the inside of the sticks measure the potential between the two sides. Resistance is determined by applying Ohm's law.
This direct current method measures the resistances of the medium, the filter, and the cell layer in sum. 
The course of the input and output signal is described in \ref{fig:messaufbau} A) and an equivalent electrical circuit is depicted in D). Cell layer resistance can be obtained by subtracting the resistance of a sample without cells from that of a sample with cells. The result is highly susceptible to wrong positioning of the electrode. Therefore, careful positioning of the electrodes is required \citep{srinivasan2015teer}. 

\subsection{Impedance}

By changing the input signal from direct to alternating currents with varying frequencies, many limitations of direct current resistance methods can be overcome.
Ohm's law defines the quotient of voltage and current as resistance. Analogous to Ohm's law for direct current,  impedance $Z$ is defined as
$Z =U_{t}/I_{t}$ as the time-dependent quotient of voltage $U_{t}$ and current $I_{t}$ for alternating current. Therefore, the frequency-dependent electrical resistance of cell barriers becomes accessible.
Current and voltage characteristics can be described by  sinusoidal functions: 

\begin{align}
U_{t} &=  U_{0}  \sin(\omega t + \Phi)  \qquad \qquad I_{t} =  I_{0} \sin(\omega t).\label{equ:Voltage}
\end{align}
Where $U_{0}$ and $I_{0}$ are the amplitudes, $\omega$ is the radial frequency, and $\Phi$ is the phase shift by which the current lags the voltage.

The impedance has a magnitude of $Z_{0} = U_{0}/I_{0}$ and by Euler's relationship, the impedance can be expressed as a complex number.

\begin{equation}
 Z=\dfrac{U(t)}{I(t)} =Z_{0} e^{i \Phi} = Z_{0}(\cos \Phi + j \sin \Phi) = Z' + j Z'' \label{equ:Complex_impedance1}
\end{equation} Hence, the impedance can be expressed as a complex number where $j = \surd(-1)$, $Z'$ is the real part, and $Z''$ is the imaginary part of the impedance. 
The real part of the im\-pe\-dance represents the resistance to current flow and the im\-aginary part of the capability of a system to store electrical energy.

Impedance data measured at a range of frequencies can be used to derive the physical and structural properties of a cell barrier by representing them with equivalent circuit models. Combining several different equivalent circuit elements in parallel or series forms such a model. Two standard circuit elements are predominantly used for modeling. The first one is the frequency-independent resistance ($Z_{R}=R$). The second equivalent circuit element is an ideal capacitor, whose impedance is frequency-dependent and can be expressed as  $Z_{C}=1/j \omega C$, where $C$ is the capacitance.

The basic elements can be combined to fit the impedance response of an electrochemical system. According to Kirchhoff's law, the total impedance of two impedances that are connected in series can be calculated by the sum of the individual impedances ($Z =  Z_{1} + Z_{2}$). In case two impedance elements are set in parallel, the inverse of the total impedance is the sum of the inverse of the individual impedances \newline (${1}/{Z}= {1}/{Z_{1}}+{1}/{Z_{2}}$) \citep{orazem2017electrochemical}.

\subsection{Cell barrier impedance}

A cell barrier posseses a capacitive and resistive character. The capacitive effects rise from the impermeability of charged species through the phospholipid membrane of cells. The conductivity of the cell membrane is $10^7$ times lower compared with the extracellular medium \citep{kotnik2000analytical}. 
When single cells start to connect through tight junctions, they control the ionic flux across the intercellular space, and form a barrier. The capacitance of a cell membrane is approximately 1 $\mu$F/cm$^2$ \citep{cole1968membranes} but can appear much larger when the cell membrane is folded \citep{lo1995impedance}. In addition, the composition \citep{wegener2004automated}  and the cell layer thickness \citep{van1999transepithelial}  influence the capacity. The biological meaning of the cell layer capacitance can, therefore, be related to morphological characteristics such as microvilli formation or other membrane alterations \citep{czupalla2014vitro}.
At high frequencies ($>$10 kHz), the impedance of the cell layer reaches zero, because of the capacitively coupled pathway through the cell barrier. The impedance at high frequencies only results from the media and filter resistance.

High electrical resistance represents a strong, intact barrier that controls the flux of substances \citep{groeber2015impedance}. 
If strongly polarized cells are measured, the surface area and protein concentration may differ from the apical to basolateral sides, which leads to different capacitances for the apical and basolateral cell membrane. The resistances and capacitances can be used to draw conclusions about biological processes in and across the barrier.

For accessing the electrical impedance, a small electrical potential or current across the cell layer must be created. Therefore, at least two electrically conductive electrodes on each side of the cell barrier are needed. In a frequency range  between 10~Hz and 1~MHz, the impedance of the media and cell support is frequency independent. However, the electrode-electrolyte interface forms a frequency dependent electrode polarization impedance, which disturbs a cell barrier measurement. Therefore, the electrode-electrolyte impe\-dance
contribution must be considered in the equivalent electric circuit modeling, or the current-carrying electrode must be separated from the potential sensing electrode. The latter erases the impedance contribution of the
current-carrying electrodes and therefore simplify the equivalent electric circuit \citep{schwan1966alternating}. 
 The commercially available system for measuring EIS in transwell  inserts CellZscope\textsuperscript{\textregistered} 
 \newline (nanoAnalytics, M\"unster, Germany) uses two fixed steel electrodes and considers the contribution of the electrode surface. Other groups integrated the four-electrode EIS setup via fixed wires \citep{onnela2012electric} or as recently presented by Nikulin et al. an approach where a chopstick-like electrode is connected to an impedance generator \citep{nikulin2019application}. 
In this work we demonstrate the powerful combination of a chopstick-like electrode and EIS in three sets of experiments. In all experiments the colon carcinoma cell line (Caco-2) was used, which reassembles the small intestine barrier \citep{hidalgo1989characterization}:
\begin{enumerate}
\item  We compared the standard epithelial voltohm method for TEER measurement with the EIS method by monitoring the development of a cell layer. The electrical impedance of a Caco-2 barrier, which was cultivated in different coated transwell inserts, was analyzed over twelve consecutive days and subsequently the capacitance and resistance extracted by electrical circuit modeling.
\item  The response of a fully developed cell barrier to different chemicals was monitored with the EIS, and the cell layer resistance and capacitance were used to characterize the effects of the substances. The complex biochemical processes of the Caco-2 layer after the addition of forskolin to the media were monitored. Fors\-ko\-lin is known to shorten the brush border microvilli \citep{rousset1985enterocytic} and increase the apical $HCO_{3}^{-}$ secretion \citep{laohapitakworn2011parathyroid} among other complex biochemical reactions related to a high concentration of the second messenger cyclic adenosine mono\-phosphate (cAMP).
\item  The effect of sonoporation was monitored by EIS. During sonoporation, ultrasound is used to oscillate or destroy gas-filled microbubbles, which temporarily increase cell membrane permeability, and thus is a pro\-mising tool to transport substances across barriers \citep{snipstad2018sonopermeation}.   The penetration of cell barriers can be investigated via the impedance-derived data TEER and capacitance \citep{kooiman2009increasing,lelu2016primary,stewart2018prototype}. EIS can be performed during the sonoporation procedure, which gives online information of the cell barrier disintegration. In long term experiments the recovery of the sonoporation of the cell barrier can be measured by EIS as well. 
\end{enumerate}

\section{Materials and methods}

\subsection{Cultivation}
\label{cultivation}
Caco-2 cell line, medium components, collagen~I, Matri\-gel\textsuperscript{\textregistered}, and TEER enhancers were all purchased from Sigma Aldrich. Eagle's minimum essential medium (EMEM) was prepared according to the manufacturer, and supplemented with 2~mM glutamine,  1\% non-essential amino acids, 1\% pen\-icillin/strepto\-mycin, and 10\% fetal bovine serum. For the reported experiments, cell passage 71 was used. The cells were seeded with a cell density of $10^5$ cells/cm$^2$ in 24-trans\-well filter (Polyester, 0.4~$\mu$m pore size, Corning Inc., Corning, NY,~USA).
The day before seeding, the inserts were coated with collagen~I  or Matrigel\textsuperscript{\textregistered} for cell adhesion and growth. Final concentrations of 50~$\mu$g/mL for collagen~I and 250~$\mu$g/mL for Matrigel\textsuperscript{\textregistered} were prepared with sterile demineralized water. A total of 50~$\mu$L of the corresponding diluted solution was applied on each insert and incubated at 37~$^\circ$C for at least 4~h. Afterward, the coating solution was discarded, and the inserts were kept in EMEM medium at 37~$^\circ$C until cell seeding. The cells were seeded and cultivated in 0.2~ml of cell culture medium with 1~ml of the same medium in the basolateral compartment and incubated at 37~$^\circ$C and 5\% CO\textsubscript{2}. The medium was exchanged every day.

\subsection{Chopstick-like electrode preparation}

Removing the RJ-11 plug and connecting the four wires from the chopstick-like electrode STX-1 (Merck, Darmstadt, Germany) to a Versastat~3F potentiostat galvanostat (AMETEK, Berwyn, PA, USA) upgrades the chopstick-like electrode to a four-electrode electrical impedance spectroscopy setup. 
On the inside of each tip, a silver/silver chloride pellet measures the voltage and is connected to the working sense and reference electrode of the potentiostat. 
The outer electrodes were connected to the working and counter electrodes. If not in use, the chopstick-like electrode was stored in phosphor buffered saline (PBS) with the two reference electrodes in short circuit to allow the electrode pair to equilibrate and limit the potential difference. The working procedure regarding cell culture experiments was equal to the standard TEER measurement with chopstick-like electrodes in transwell inserts and was described by the manufacturer of chopstick-like electrodes. In short, the electrodes were sterilized with 70\% ethanol and dipped afterward in fresh cell culture media to remove the excess ethanol. The cells grown in transwell inserts were taken out of the incubator and placed in a sterile workbench. The chopstick-like electrodes were immersed in the transwell with the short tip on the inside and the long tip outside of the insert. 
 
 The EIS measurements were performed in galvanostatic mode with an amplitude of 10~$\mu$A and at a frequency range of 10~Hz to 100~kHz with ten probes per decade. Repeated EIS measurements showed no influence on the impedance response of the cell layer, which led to the conclusion that the designed EIS measurement is a noninvasive method and that the cells were not harmed by the measurement. Moreover, no difference in the impedance response was found by changing between potentiostatic and galvanostatic modes. 
The impedance spectra were analyzed with the fitting software Zview~2 (Version 3.5b, Scribner Associates Inc., SP, USA). The data were fitted with an equivalent electric circuit, as shown in Figure \ref{fig:messaufbau} E), using a nonlinear least-squares fitting algorithm.  

To ensure comparability between the TEER values generated by EIS measurement and standard voltohm meters (e.g. EVOM, Millicell) a fully developed cell layer was measured with both. The resulting TEER was similar. For further comparison of our proposed method to the voltohm resistance measurement over different stages of cell barrier formation, we adopted the input signal from voltohm devices. According to the manufacturer, a direct current of 10~$\mu$A with 12.5 polarity reversal per second is used as the input signal. By chronopotentiometry measurements with a direct current of 10~$\mu$A or -10~$\mu$A alternating for 80~ms, the voltohm settings can be mimicked with a potentiostat. The measurement was canceled after ten positive and negative currents. At the end of each cycle, the potential was recorded, and the resistance was calculated. The means of all twenty resistances were the cell layer resistances after subtracting the resistances of an empty transwell. 
The formation of the epithelial cell layer was monitored daily by EIS and the above described voltohm method. 

\subsection{EIS response to TEER enhancer}
The impedance response of a fully developed Caco-2 cell barrier to three different TEER enhancers, namely, EGTA, saponin and forskolin was recorded.
The stock solutions of EGTA and saponin were prepared in culture medium. The final concentrations were chosen according to literature, hence 0.05~\% w/v \citep{chao1998enhancement} 
for saponin and 10 mM \citep{sun2010chip} for EGTA. 10~mg forskolin was first dissolved in 1~ml ethanol (=24 mmol/l), and 
18.5 $\mu$L of the forskolin solution in ethanol was added to 45 ml medium leading to a 10 $\mu$M solution~\citep{rousset1985enterocytic}.  The chopstick-like electrode was placed in a transwell insert. Prior to the experiment, a full EIS spectrum was recorded to determine the starting value for TEER and cell layer capacitance. All the inserts showed at least a starting TEER of 1500 $\Omega $ cm$^{2}$ and a total membrane capacitance of 3  $\mu$F/cm$^{2}$. After an EIS spectrum, the media on the apical side were replaced with medium spiked with a TEER enhancer and EIS was performed in a loop to monitor the immediate response of the Caco-2 cell layer to the TEER enhancer. The measurement was terminated when the alteration between the loops was negligible. The values for TEER and cell layer capacitance were extracted by equivalent electric circuit fitting.

\subsection{Sonoporation}
\label{sonoporation}

For analyzing the influence of sonoporation to a Caco-2 cell layer, the impedance response was measured during the application. For the experiments, microbubbles with a polymeric n-butyl cyanoacrylate shell were used. The used microbubbles are described elsewhere \citep{appold2017physicochemical}.
A 1 MHz transducer (Olympus, Waltham, MA) was placed in a water bath and the transwell plate was mounted closely under the
water surface. Ultrasound was applied with a pressure of 600 kPa for three min.
The microbubbles were freshly mixed before each sonoporation experiment to a concentration of 1.8 Mio bubbles per ml medium and injected to the lower compartment of the well plate as the microbubbles float. The Caco-2 cells were cultivated on the lower side of an insert for direct contact between the cells and microbubbles.

\subsection{Fluorescence microscopy}
\label{fluorescence}

The transwell inserts were rinsed with cooled PBS buffer and fixated by 4\% paraformaldehyde in PBS for 15~min. Followed by three washing steps with PBS containing 5\% bovine serum albumin (BSA) and 5\% fetal calf serum for 30 min at room temperature. The cells were then washed three times for 15 min with PBS containing 1\% BSA, and the inserts were incubated overnight at 4~$^\circ$C with the first antibody rabbit anti-ZO-1 (Thermo Fisher Scientific Inc., Waltham, MA, USA) which was diluted in PBS with 1\% BSA to a concentration of 2.5~$\mu$g/mL. After decanting the solution and washing three times for 15 min in PBS with 1 \% BSA the second antibody goat anti-rabbit with Alexa-488 (Thermo Fisher Scientific Inc., Waltham, MA, USA) with a concentration of 5~$\mu$g/mL was applied to the cells for 60 min at room temperature in the dark. Lastly, after one washing step in PBS with 1~\% BSA and two washing steps in PBS without BSA for 15 min, the cells were incubated with 100~$\mu$g/mL DAPI (4',6-diamidino-2-phenylindole) diluted in PBS for 5 min in the dark and at the end washed with PBS. The images were taken with the  Leica SP8 and ZEISS ApoTome confocal laser scanning microscopes.

\section{Results and discussions}
The first part of this section explains the principles followed for the development of an electrical equivalent circuit model which is capable of describing the impedance response of a cell layer. Subsequently, the results of different experiments show the versatility of the model. 

\subsection {Development of the equivalent electrical circuit}

The equivalent electrical circuit should fulfill the following criteria:

\begin{itemize}
\item The impedance response of a cell layer between 10 Hz and 100 kHz should be described at all stages of maturation and additionally detect complex biological processes.

\item The amount of fitting parameters must be kept to a minimum in order to achieve a proper physical representation of single elements.

\item The cell layer and the measurement path must be reflected.

\item A direct comparison with the TEER measured with a chopstick-like electrode connected to a voltohm device (DC-measurement) should be possible, since this is the most widespread method to characterize barrier-forming cells. 
\end{itemize}

Using a four-electrode EIS setup has the benefit of eliminating electrode polarization resistance from the impedance spectra, leading to a frequency-independent measurement setup. Therefore, without cells, the impedance response is purely resistive and arises from the cell culture media, membrane support and wires. These resistances are summarized in a single resistor element $R_{media}$. 

An equivalent electrical circuit that describes a cell layer needs at least a resistive and a capacitive element in parallel. Resistance TEER describes the flux of ions through the barrier. The capacitance represents the capacitive behavior of the phospholipid membrane of the cells. If the cell layer is polarized, which means that the cell membrane of the apical and basolateral sides differs in surface area and protein composition, two capacitances must be considered.  

\begin{figure}
	\centering
		\includegraphics[width=1\columnwidth]{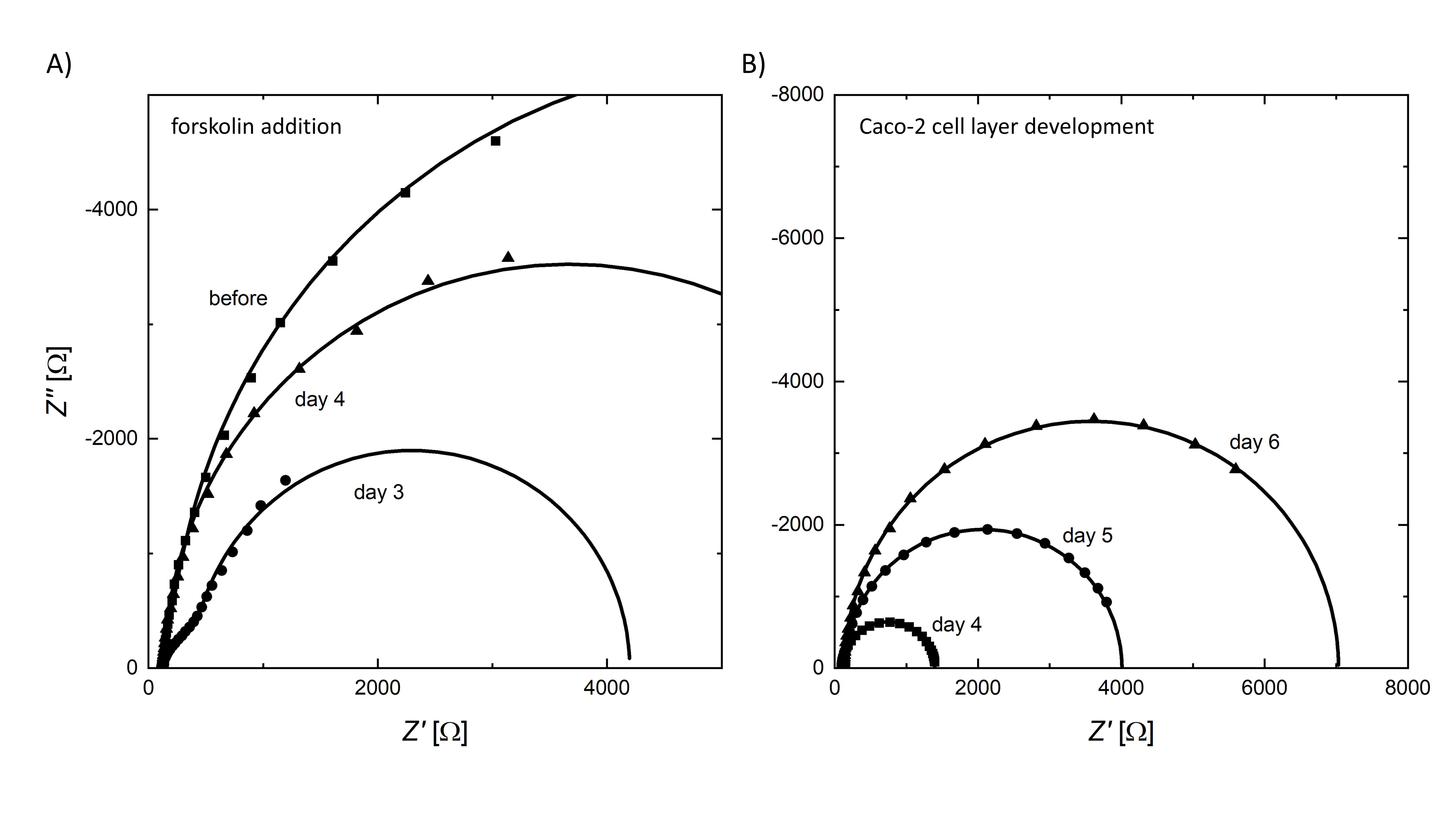}
	\caption{Nyquist plots of the EIS measurements of a Caco-2 layer between 10~Hz and 30~kHz. The data were fitted with the equivalent circuit presented in Figure \ref{fig:messaufbau} E): A) before the addition of forskolin to the cell culture media and three and four days after addition respectively. Three days after the forskolin addition two semicircles appear. B) The development of the Caco-2 barrier at three different days showing good agreement between measurement and fitting.}
\label{fig:nyquist}
\end{figure}

 The polarized Caco-2 cell layer was fitted with a resistance and a capacitance in parallel for the apical and another resistance and capacitance for the basolateral cell membrane. Figure \ref{fig:messaufbau} E) presents the complete equivalent electrical circuit. This circuit allows for fitting complex biological processes as demonstrated by the addition of forskolin, as well as leaky and tight epithelia, as shown in Figure \ref{fig:nyquist}. Forskolin alters the electrochemical behavior of the membrane surface by shortening the brush border microvilli \citep{rousset1985enterocytic} and increasing the apical $HCO_{3}^{-}$ secretion \citep{laohapitakworn2011parathyroid} among other complex biochemical reactions related to a high concentration second messenger concentration cAMP. Figure \ref{fig:nyquist} A) shows first a section of the nyquist plot for the impedance measurement before the addition of forskolin. After day 3 of forskolin addition, two semicircles appear in the Nyquist plot. The two semicircles appear when the time constant ($\tau$ = R*C) of the two RC elements differ sufficiently. Figure \ref{fig:nyquist} B) shows three exemplary nyquist plots during the development of the barrier and the corresponding fit.

 The equation for the used equivalent electrical circuit is depicted here:
 
\begin{align}
Z &=  R_{media} + \dfrac{R_{1}}{1+j\omega R_{1} C_{1}} +\dfrac{R_{2}}{1+j\omega R_{2} C_{2}} 
\label{equ:impedanceresponse}
\end{align} 
  An additional resistor is often used parallel to the apical and basolateral resistor and capacitance to differentiate between paracellular and transcellular transports. The increased number of undefined parameters must be solved by further simplifying assumptions or specialized experiments  \citep{lewis1984apical,krug2009two}. 
Summing up, for measurements with polarized cells, two resistance and capacitance elements are necessary. However, considering  apical and basolateral resistances and capacitances separately is often impractical. 
Therefore, the derived resistances and capacitances may be summarized according to Eq. (\ref{equ:Teerandcap}) in TEER and total membrane capacitance C as reported in literature \citep{erlij1994effect,bertrand1998system}.

\begin{align}
TEER &=  R_{1} +  R_{2} \qquad \qquad 
\dfrac{1}{C} = \dfrac{1}{ C_{1}} +\dfrac{1}{ C_{2}}
\label{equ:Teerandcap}
\end{align} 
These assumptions enable the comparison with experiments performed with the commercial EIS device for transwell inserts CellZscope\textsuperscript{\textregistered} or TEER data from chopstick-like electrodes connected to an EVOM or Millicell voltohmmeter. 

\begin{figure*}

		\includegraphics[width=2\columnwidth]{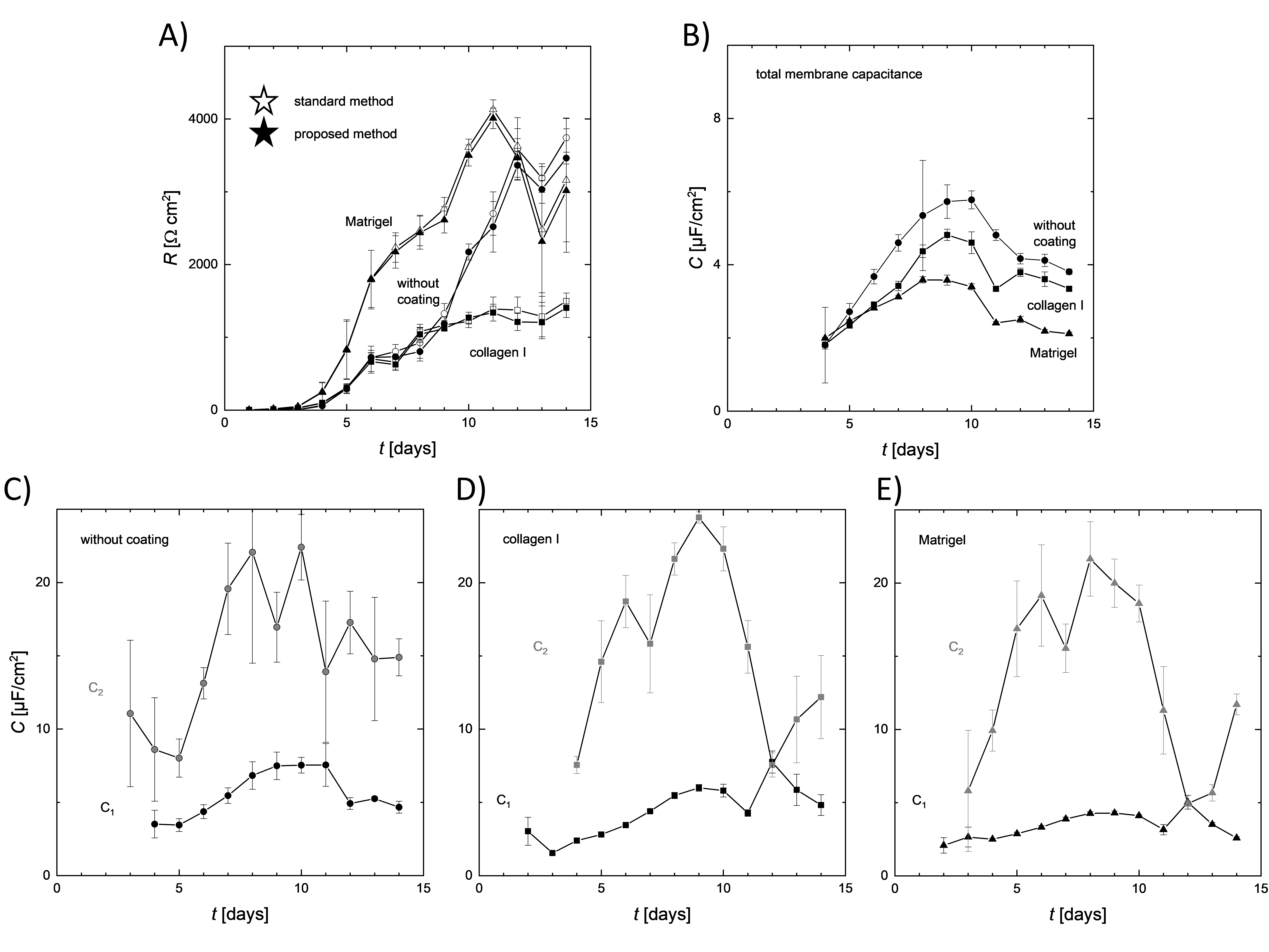}
	\caption{The effect of membrane coating on the development of a Caco-2 cell barrier was studied with respect to resistance and capacitance measurements over a time period of 14 days. The inserts were coated with Matrigel\textsuperscript{\textregistered} and collagen~I and compared with inserts without coating. The standard voltohm settings (10 $\mu$A direct current with 12.5 polarity reversal per second) were used to verify the resistance measured with the proposed chopstick-like electrode impedance method. A) shows the development of the TEER measured with both methods and B) shows the corresponding overall cell layer capacitance. The three lower graphs show the development of capacitances $C_1$ and $C_2$ separately. The measurements were done at least in triplicate and capacitances higher 25 $\mu F/cm^2$ were neglected.
	}
\label{fig:teerevom}
\end{figure*}

\subsection{Caco-2 cell layer development}

The most common devices for TEER measurement in transwell inserts are voltohmmeters. These devices apply a 10~$\mu$A direct current with polarization reversal every 80~ms. 
We adopted the voltohm settings to compare the TEER data derived from voltohm with the TEER data derived from EIS. The voltohm and EIS settings were implemented as a sequence to measure consecutively without changing the chop\-stick-like electrode position. The development of the TEER on different coated inserts was monitored over two weeks and is shown in Figure \ref{fig:teerevom} A). The TEER values derived from voltohm or EIS settings are in good agreement during all the stages of barrier maturation. The cells grown on Matrigel\textsuperscript{\textregistered}-coated inserts developed the fastest and reached roughly the same end value as the inserts without coating. The cells grown on collagen-coated inserts developed the same TEER as the cells grown without previous coatings until day 9 and stayed lower after day 9 to around half of the TEER of Ma\-tri\-gel\textsuperscript{\textregistered} or without coating.  

The development of cell layer capacitance shows a different course over time. The corresponding total membrane capacitance of the cell layer can be seen in Figure \ref{fig:teerevom} B) and is only accessible through the proposed EIS method. While the total membrane capacitance shows a similar trend, the two capacitances reveal differences in the temporal development. The total membrane capacitance is dominated by capacitance $C_1$, which is reported to correlate to the apical cell membrane. The basolateral side is represented by $C_2$ \citep{schifferdecker1978ac, bertrand1998system}. The membrane area of the basolateral side is larger than the apical side because the intercellular space is included \citep{clausen1979impedance}.  The apical and basolateral capacitance are separately presented  in Figure \ref{fig:teerevom} C) - E). 

The coating had a strong effect on the capacitance of the cell layer. All the cell layers peaked between days 8 and 10. The lowest peak total membrane capacitance was shown by the cells grown on Matrigel\textsuperscript{\textregistered} with approximately 3~$\mu F/cm^2$ compared with almost 6~$\mu F/cm^2$ for the uncoated inserts. 
The data interpretation for the capacitance development over the 14 days of cultivation is challenging. Leffers et al. reported comparable TEER and capacitance values with 
 1500 $\Omega cm^2$ and $4~\mu F/cm^2$, respectively, for collagen-coated transwell inserts \citep{leffers2013vitro}. The capacitance correlates with the cell membrane area and protein composition. Further research is necessary to correlate the capacity of the cell layer with biological processes, such as the degree of polarization.

\subsection{Impedance response to TEER enhancers and forskolin}

\begin{figure}[pos=h]
	\centering
		\includegraphics[width=1\columnwidth]{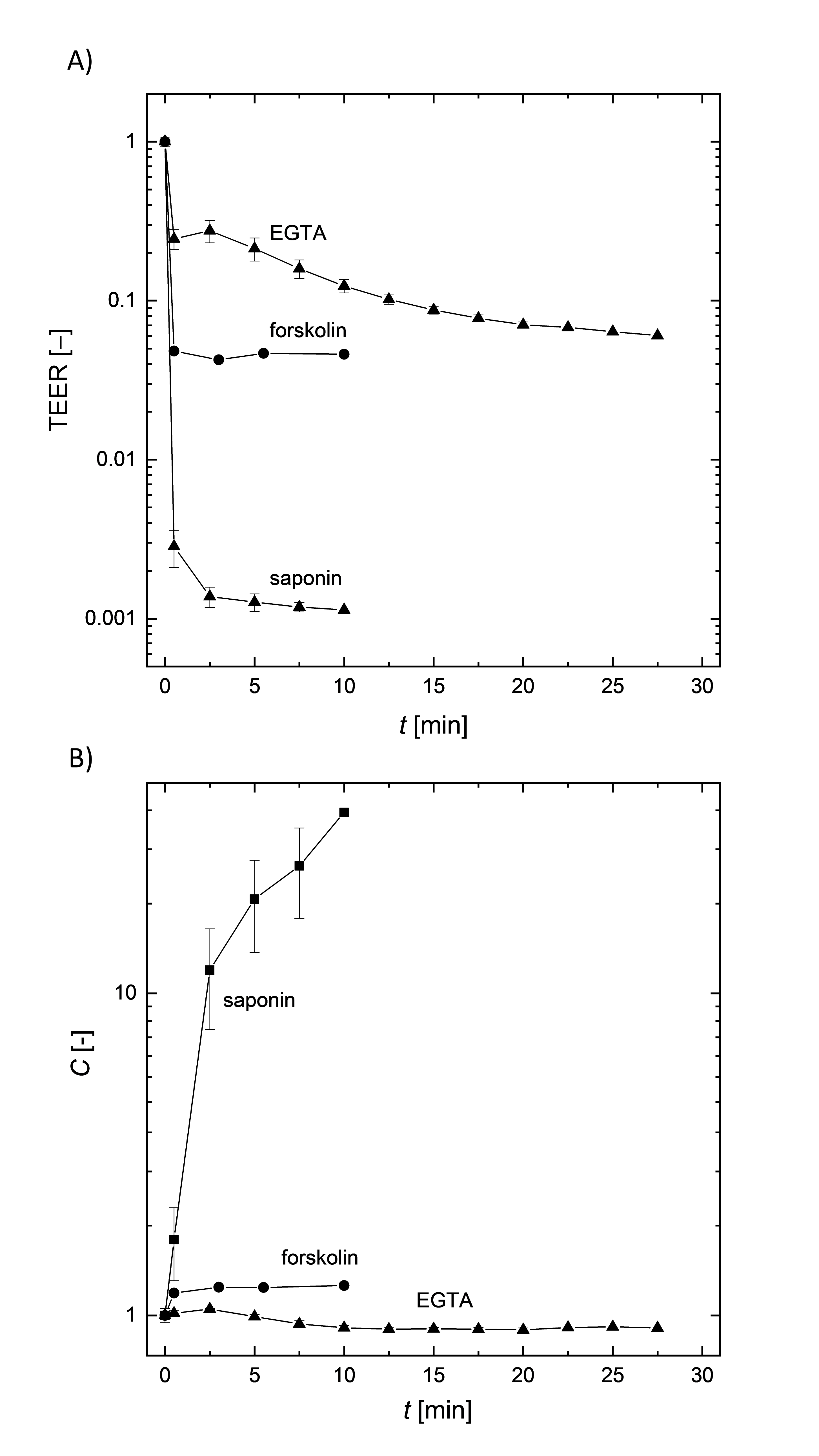}
	\caption{The graphs show the immediate TEER (A) and capacitance (B) response of Caco-2 cell layers to EGTA, forskolin, and saponin. The values were normalized to the beginning TEER and capacitance value. All three chemicals sharply decrease the TEER, but only saponin alters the cell layer capacitance significantly. Data shown represent the mean of three cell monolayers.
	}
	
\label{fig:teerenhancer}
\end{figure}

The response of a Caco-2 barrier to EGTA, saponin, and forskolin was assessed by measuring the impedance spectra with a chopstick-like electrode. The TEER and capacitance were extracted by fitting the impedance data to the equivalent circuit depicted in Figure \ref{fig:messaufbau} E). The resistances and capacitances were summarized according to Eq.~(\ref{equ:Teerandcap}) and normalized to the initial values and presented in Figure~\ref{fig:teerenhancer}.
While all three substances decreased the TEER value substantially (\ref{fig:teerenhancer} A), only saponin altered the capacitance (\ref{fig:teerenhancer} B) significantly, due to the hemolytic effect of this non-ionic surfactant \citep{narai1997rapid}. The tremendous increase in capacitance is correlated to the existence of apoptotic or detached cells \citep{lohren2016effects}.  The calcium-chelating agent EGTA reduced the TEER within the first 20~min by approximately 90~\%, but the capacitance remained constant, thereby leading to the assumption that only the tight junctions are affected by EGTA. Forskolin with its complex biological effects resulted in an instant drop in the resistance of nearly 95~\% within the first 5 min. Analogous to the experiments with EGTA, the capacitance stayed almost constant with respect to the initial value, which leads to the assumption that forskolin has an immediate effect on the cell\textendash cell connection but no effect on the morphology as this would alter the capacitance.

\subsection{Fluorescence microscopy}

\begin{figure*}
		\includegraphics[width=2.0\columnwidth]{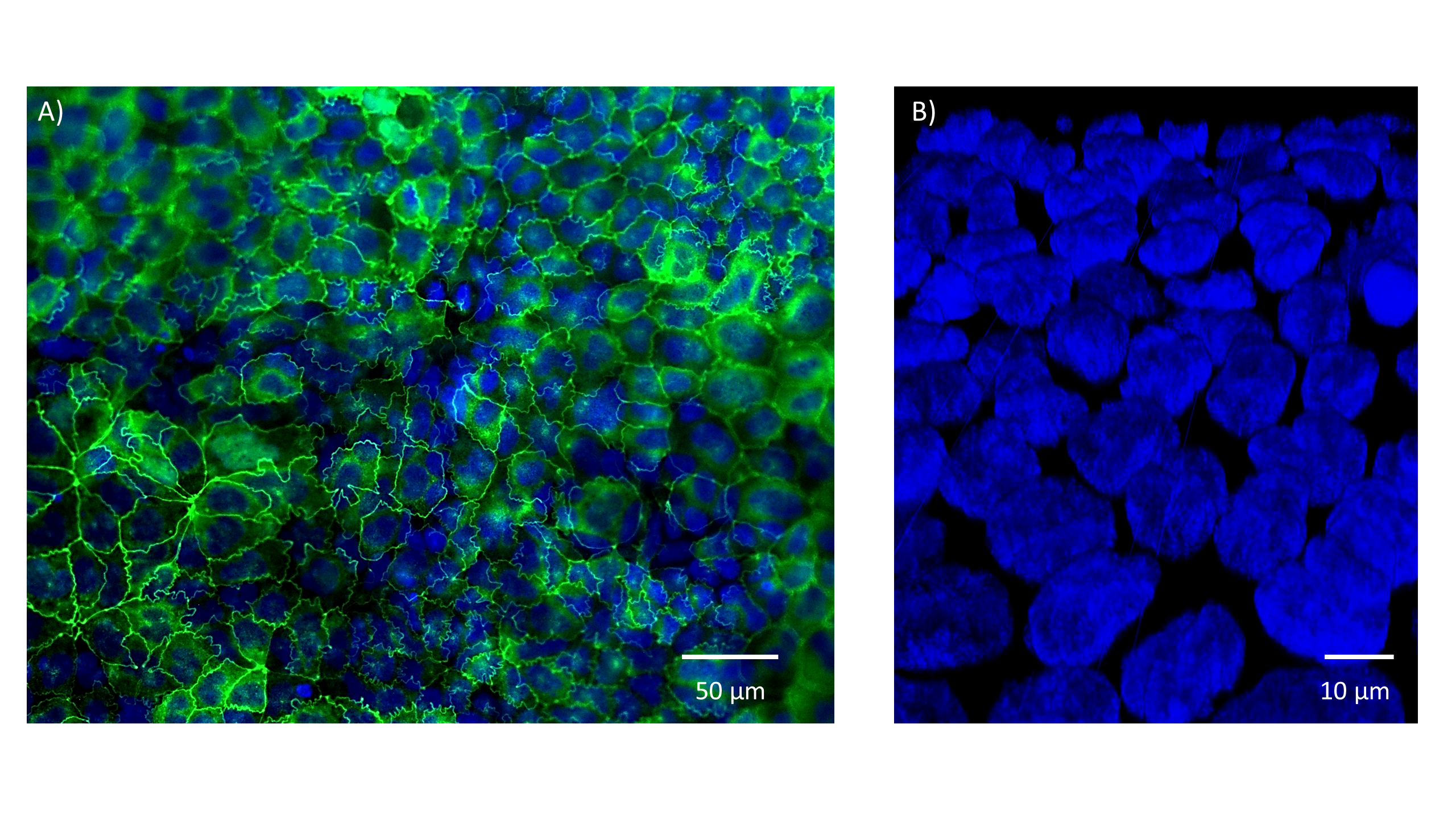}
	\caption{A) Fluorescence microscopy revealed a dense and homogenous Caco-2 epithelial cell layer. In addition, z-scans showed that the cells were grown in monolayer B). Blue: DAPI staining of the cell nucleus, green: ZO-1 antibody staining. Cells grown on collagen~I or Matrigel\textsuperscript{\textregistered} coated membranes showed no differences in fluorescence microscopy.}
\label{fig:zo1}
\end{figure*}

The staining of ZO-1 protein, which is part of the tight-junction complex, and the nucleus revealed the cell morphology in Figure \ref{fig:zo1}. A confluent cell layer with distinct tight junctions was visualized by fluorescence microscopy, and the Caco-2 cells were grown in monolayer according to the confocal microscopy images. Unfortunately, no difference between cells grown on different coated membranes can be observed. 

\FloatBarrier

\subsection{Sonoporation}

\begin{figure}[pos=h]
	\centering
		\includegraphics[width=0.9\columnwidth]{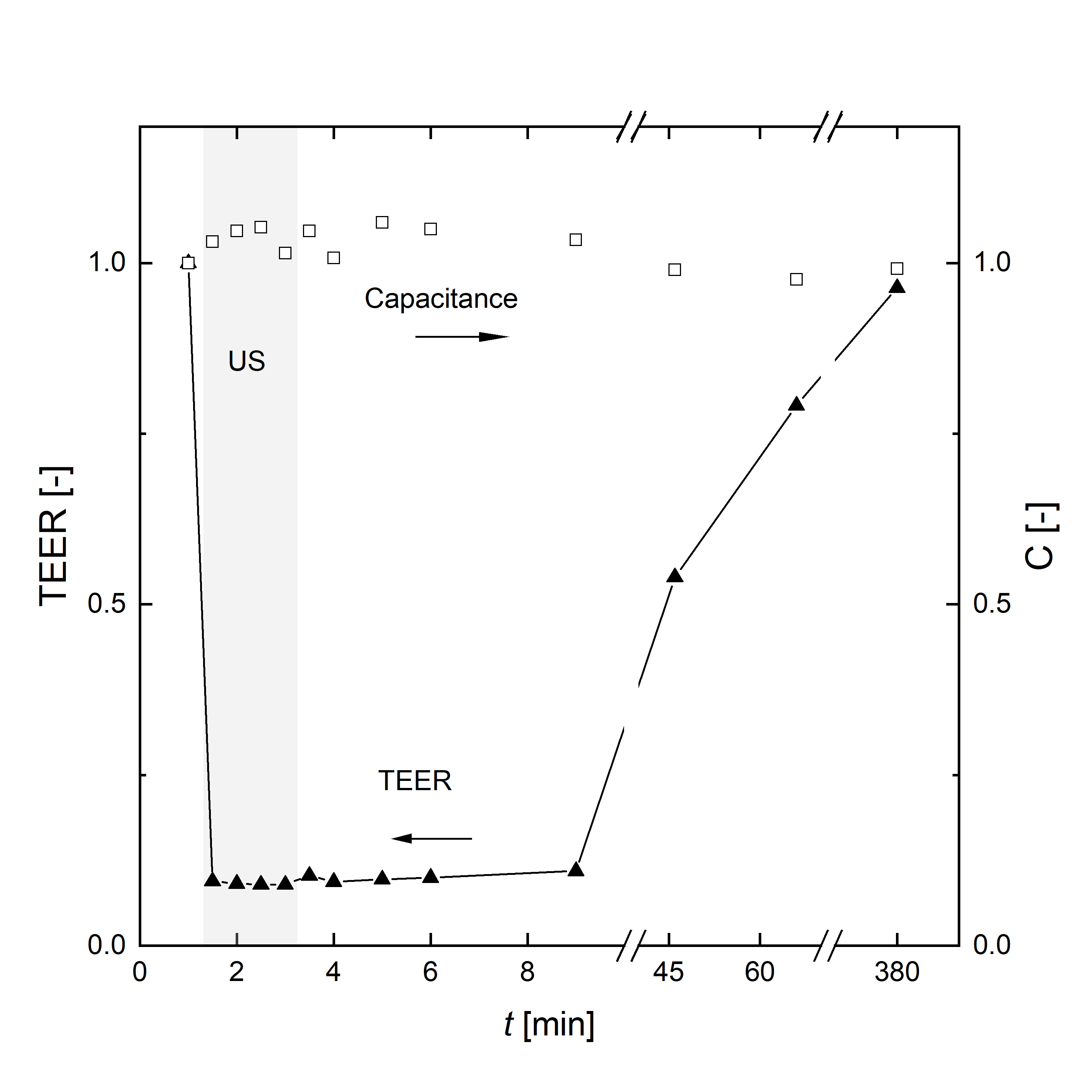}
	\caption{ 
	Normalized TEER and total membrane capacitance during and after sonoporation. The gray area shows the time when the ultrasound (US) was applied. The resistance was reduced directly with the start of the sonoporation and increased to the initial value again within 6 hours. The total membrane capacitance remained almost constant.
	}
\label{fig:sonopermeation}
\end{figure}

Figure \ref{fig:sonopermeation} shows the course of the TEER and total membrane capacitance during a sonoporation of a Caco-2 cell layer. By applying ultrasound in combination with microbubbles in contact with the cell layer the TEER instantly dropped to less than 10\% of the initial value. During the sonoporation the TEER stayed constant. 45~min after turning off the ultrasound, the TEER increased to  54\%. After nearly 6.5~h, the initial TEER was almost reached with a percentage of 99\%. The capacitance stayed almost constant and fluctuated in a range of 4\% around the initial value. No change in the impedance signal was observed when using microbubbles or ultrasound alone. 

\section{Conclusions}

The use of a chopstick-like electrode is the most convenient way to measure TEER from cells grown in transwell inserts. The TEER represents the transport of ionic species through the barrier and can be used to measure the cell layer confluence and monitor the experiments related to ionic transport. 

By changing the input signal from direct to alternating currents with varying frequencies, the frequency-dependent resistance (impedance) can be calculated. By fitting the data to an equivalent electric circuit, the media resistance, cell layer resistance, and cell layer capacitance can be extracted. The cell layer capacitance correlates to the surface area and composition of the cell membrane. 
In a first step, we showed that the TEERs derived from the standard and the introduced method are comparable during all the stages of barrier formation. The additional cell layer capacitance can be used to further monitor the morphological behavior of the cells, e.g., the development of a polarized cell layer.
The proposed method was then used to monitor the behavior of fully developed Caco-2 barriers to the permeability enhancers EGTA and saponin. Both decreased the TEER substantially, but only saponin altered the capacitance. Without the proposed impedance method the differentiation between the two substances would not be possible. 
A crucial step in impedance spectroscopy is to use an appropriate equivalent electrical circuit. The used circuit allows the fitting of the experimental data during all the stages of barrier formation and during complex biological events, such as after the addition of forskolin. The number of fitting parameters is summarized to gain two easy-to-handle values, TEER, and total membrane capacitance. 
In the last set of experiments, sonoporation was monitored. The TEER drops immediately while the capacitance remains constant during ultrasound and microbubbles exposure. 

Our introduced method can help barrier researchers to monitor and evaluate their experiments by the additional capacitance data without changing the established workflows. 
 
\section*{Conflict of interest statement}
The authors declare that there is no conflict of interest. 
\section* {CRediT authorship contribution statement}
\textbf{G. Linz:} Conceptualization, Methodology, Formal analysis, Investigation, Data Curation, Writing - original draft, Writing - review \& editing, Visualization, Supervision.
\textbf{S. Djeljadini:} Investigation.
\textbf{L. Steinbeck:} Investigation, Visualization.
\textbf{K. Gurbet:} Investigation.
\textbf{F. Kiessling:} Resources, Writing - review \& editing.
\textbf{M. Wessling:} Conceptualization, Resources, Supervision, Methodology, Funding acquisition, Writing - original draft, Writing - review \& editing.

\section*{Acknowledgements}
M.W. acknowledges the support through the Gottfried Wilhelm Leibniz award. This project has received funding from the European Research Council (ERC) under the grant agreement no. 694946. Part of the work was performed at the Center for Chemical Polymer Technology CPT, which is supported by the EU and the federal state of North Rhine-Westphalia (grant no. EFRE 30 00 883 02).

\bibliographystyle{cas-model2-names}

\bibliography{cas-refs}

\end{document}